# 3D modeling of magnetic atom traps on type-II superconductor chips

Vladimir Sokolovsky[1], Leonid Prigozhin[2], John W Barrett[3]

[1]Physics Department, Ben-Gurion University of the Negev, Beer-Sheva, 84105 Israel
[2]A. Yersin Department of Solar Energy and Environmental Physics
The Blaustein Institutes for Desert Research
Ben-Gurion University of the Negev, Sede Boqer Campus, 84990 Israel
[3]Department of Mathematics, Imperial College London, London SW7 2AZ, UK.

**Abstract**
Magnetic traps for cold atoms have become a powerful tool of cold atom physics and condense matter research. The traps on superconducting chips allow one to increase the trapped atom life- and coherence time by decreasing the thermal noise by several orders of magnitude compared to that of the typical normal-metal conductors. A thin superconducting film in the mixed state is, usually, the main element of such a chip.

Using a finite element method to analyze thin film magnetization and transport current in type-II superconductivity, we study magnetic traps recently employed in experiments. The proposed approach allows us to predict important characteristics of the magnetic traps (their depth, shape, distance from the chip surface, etc.) necessary when designing magnetic traps in cold atom experiments.

## 1. Introduction

Laser cooling and atom trapping techniques, supplemented by technological developments in lithography and nanofabrication, made atom chips (conducting microstructures on planar substrates) a valuable tool of a new exciting area of research where solid-state, atom and light physics meet. Magnetic traps for cold atoms based on these chips are useful for studying atom-surface interactions (e.g. the Casimir-Polder force), the spin decoherence of atoms near dielectric bodies, and in the usage of trapped atoms to probe local irregularities of magnetic and electric fields near conductive films. Possible applications of atom chips are in quantum information processing, quantum metrology, quantum optics, high-resolution spectroscopy, atom SQUIDs and interferometers, etc. [1-8].

The chips produce tiny magnetic field configurations which can trap, cool, and manipulate ensembles of ultra-cold atoms in a deep vacuum near solid surfaces [1-6]. Trapping neutral atoms having a nonzero magnetic moment is based on the Zeeman effect: depending on the quantum state of an atom with a magnetic moment, the atom's energy either increases or decreases with the magnetic field growth. In a non-uniform field, the atoms whose energy decreases with the field tend to occupy the position where the field is the strongest. Since no maximum of the magnetic field can be created in the free space (the Earnshaw theorem), the high-field-seeking atoms cannot be trapped. It is, however, possible to create a minimum of the magnetic field magnitude and trap the low-field-seekers: the atoms whose magnetic energy increases with the magnetic field.

The trapped atom kinetic energy should be much lower than the depth of the magnetic trap potential well, and modern laser cooling technology (e.g. a successive application of the Doppler cooling, Sub-Doppler cooling, etc.) allows one to achieve an atom cloud temperature of the order of tens of nano-Kelvins [5]. To ensure that the kinetic energy is smaller than the trap depth, a typical trapping criterion is [1,4,9] $\mu B_{dep} \geq 10 k_B T$, where $\mu$ is the atom magnetic moment, $k_B$ the Boltzmann constant, $T$ the atom cloud temperature, and $B_{dep}$ is the trap depth, the difference between the maximal level of the magnetic field magnitude, for which the iso-surface of $|\boldsymbol{B}|$ is closed, and the minimum of $|\boldsymbol{B}|$ inside the trap. Away from this minimum, the magnetic field gradient in the trap should be sufficient to prevent such forces as the Casimir-Polder force and gravity to pull the atoms out of the trap. At a distance from the surface larger than 1 μm the gravitational force usually dominates [1,4,9].

Adiabatic motion of the atom magnetic moment in the field is another condition for stable trapping: the moment orientation with respect to the field direction should not change. In a trap with the zero field minimum, i.e. a quadrupole trap, the atoms arriving at a low field area inside the trap easily undergo a nonadiabatic spin flip and are then expelled from the trap; this is the Majorana instability [10]. The criterion for adiabaticity is typically that the trapping frequency is smaller than the Larmor frequency. Consequently, to prevent the Majorana loss, a nonzero minimal field magnitude in a trap is desirable (as this increases the Larmor frequency). Thus, if the trap is created by a current in a long wire and a bias field perpendicular to the wire (the "side guide" configuration), an additional magnetic field is usually applied parallel to the wire [4]. Another method to decrease the Majorana atom loss is to again make a non-zero minimum, but this time with a time-averaged potential (i.e. by rotating the trap) [11].

In most of experiments, if the trap distance from the surface of a conventional conductor does not exceed a few μm, the Johnson thermal magnetic noise exceeds all other harmful influences on the atom cloud and dominantly limits its lifetime (as long as technical noise is kept to a minimum). Replacement of usual conductors by superconductors significantly decreases this noise and, according to the theoretical estimate [12], the lifetime of atoms trapped near a superconducting layer in the Meissner state can be at least six orders of magnitude longer. Analysis [13] suggests that in this case, even at the trap height of 1 μm above a superconducting layer, the cloud lifetime is limited mainly by environmental noises and may reach 5000 s, while the lifetime of an atom cloud at such a distance from a normal metal current-carrying layer would not exceed 0.1 s. Other advantages of superconducting wires are zero heat generation and the ability to carry a persistent current; the latter enables one to eliminate the current supply fluctuations and increases the lifetime.

Magnetic traps on superconductors in the Meissner state have been realized in several experiments (see, e.g., [7,14-16]). In [9,17-19] the magnetic field for such traps was estimated using the sheet current density in an infinite strip in the Meissner state. For thin films of an arbitrary shape, the distribution of the Meissner current can be found numerically, solving the London equations by a finite element method [20-22].

Atom traps with high magnetic fields, as needed for strong confinement in some applications, may not be created on chips operating in the Meissner state because magnetic vortices penetrate into the superconductor (the mixed state). Moreover, most of superconductors used in magnetic atom chips are thin superconducting films, for which partial penetration of magnetic vortices is especially difficult to avoid. Although in the presence of vortices there are magnetic field fluctuations caused by random hopping of vortices form one pinning center to another, the resulting magnetic noise is still much weaker than that near a conventional conductor (see [23-25]). Also in the mixed state the induced current can be persistent and lossless.

Contrary to the Meissner state, described by a linear model and demonstrating no memory effect, the mixed state of type-II superconductors exhibits hysteretic behavior. Type-II superconductors enable one to create a magnetic trap generated by a closed-loop persistent current [25-27] as well as by a frozen magnetic flux [28-30]. Choosing the thin film shape and applying different sequences of transport currents and/or external magnetic fields normal to the film plane, it is possible to set different stable trap configurations. Additional control of the trap depth, shape, and distance from the chip surface can be achieved by varying the bias magnetic field parallel to the superconducting film.

The distribution of current in a superconductor in the mixed state is well described by the Maxwell equations (with the displacement current omitted) supplemented by a highly nonlinear current-voltage relation; a power law relation [31] or its high power limit, the Bean critical-state model [32], are typically employed. It is usually assumed that the first critical magnetic field, $B_{c1}$, is negligibly small.

In the infinitely thin film approximation, the current-voltage law relates the electric field to the sheet current density, i.e. the current density integrated over the film thickness. This approximation is very accurate if the thickness is much smaller than the linear sizes of the film cross-section [33,34], and this condition usually holds for the superconducting films in atom chips: the typical thickness of these films is 200-900 nm while their characteristic cross-section sizes are from tens of micrometers to several millimeters.

Analytical solutions to thin film magnetization and/or transport current problems are known for the Bean model and the simple film shapes for which the current density distribution is one-dimensional (an infinite strip, disk, and ring) [35-40]. These analytical solutions have been used to analyze the magnetic trap potential; see, e.g. [9,41]. Numerically, the traps on thin films of such simple geometries were analyzed for the power law relation in [42].

For a superconducting film of a general shape, a two-dimensional problem for the distribution of current has to be solved and a variety of numerical schemes for thin film magnetization problems in type-II superconductivity have been developed [43-48]. Recently, the mixed finite element method [48] was simplified and extended to transport current problems [49]. Although this method assumes a power current-voltage relation, it remains robust for any power and, if the power is high, produces an accurate approximation to the Bean model solution. For completeness, this method is briefly described in the Appendix. Provided the distribution of the film current density is computed, the three-dimensional magnetic field, determining the magnetic trap potential, can be found numerically using the Biot-Savart law (see, e.g. [21]).

We use this numerical approach for modeling the superconducting atom chips operating in the mixed state as employed in several recent experiments. The modeling is performed in dimensionless variables: we normalize the coordinates by a characteristic film size $w$, the sheet current density $J$ by its critical value $J_c$, the transport current $I$ by the critical current value $I_c = J_c w$, and the magnetic induction $\boldsymbol{B} = \mu_0 \boldsymbol{H}$ by $\mu_0 J_c$, where $\mu_0$ is the magnetic permeability of vacuum. In the adiabatic approximation, we assume the atom cloud shape can, approximately, be represented by the shape of the closed $|\boldsymbol{B}|$ iso-surface chosen in accordance with the atom cloud temperature.

## 2. A trap on a square chip

The magnetic trap for ultracold $^{87}$Rb atoms has been created in [30] by a current in a 800 nm thick 1 mm × 1 mm square YBCO film. The film was cooled below the critical temperature $T_c$ in a zero field, and then the supercurrent was induced in it by two consequent opposite pulses of a uniform external magnetic field $B_z$, perpendicular to the film and estimated [50] as $0 \to 3\mu_0 J_c \to 0$ and $0 \to -0.8\mu_0 J_c \to 0$, where the characteristic field $B_{ch} = \mu_0 J_c \approx 100\,\mathrm{G}$; see Section 4 for a general discussion of the parameter values for different superconductor materials, temperatures, etc.

Taking a very high power ($p = 10^7$) in the power current-voltage relation $\boldsymbol{E} = E_0 (J/J_c)^{p-1} \boldsymbol{J}/J_c$ we obtain, using the method [49], a numerical solution corresponding to the Bean critical-state model (the efficiency of this method does not depend on the power value). For the finite element mesh of about six thousand elements calculating the resulting current density distribution took 11 minutes on a PC with the i5 IntelCore 3.1 GHz processor and 16 GB RAM. To compare our simulations and the experiment, in this example we return to dimensional variables.

Qualitatively, our numerical simulation (figure 1, left) agree with the experiment [30] (figure 1, right) sufficiently well: the calculated magnetic trap shape is similar to that of the atom cloud, and the calculated trap distance from the film is close to 0.16 mm (as observed in [30]). In this case the iso-surface $|\boldsymbol{B}| = 0.06\mu_0 J_c = 6$ G corresponds to stable trapping of atoms at 200 µK [30] (the temperature trap depth $\mu B_{dep}/k_B \approx 10^3$ µK).

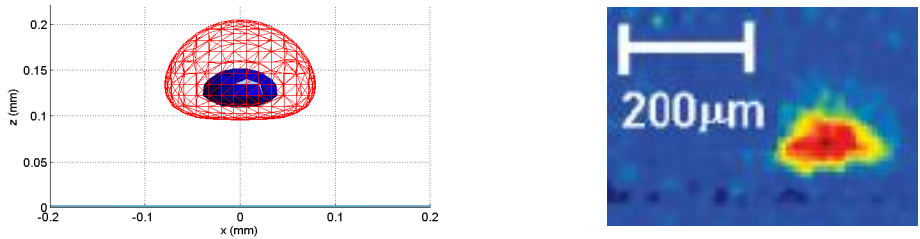

**Figure 1**. Magnetic trap on a square chip generated by two pulses of the external field $B_z$: $0 \to 3\mu_0 J_c \to 0$ and $0 \to -0.8\mu_0 J_c \to 0$. Left: computed magnetic field iso-surfaces, $|\boldsymbol{B}| = 0.03\mu_0 J_c = 3$ G (blue solid surface) and $|\boldsymbol{B}| = 0.06\mu_0 J_c = 6$ G (red lines). Right: the atom cloud image (from figure 4 in [30]).

Since the film supercurrent is induced by the magnetic field pulses $0 \to 3\mu_0 J_c \to 0$ and $0 \to -0.8\mu_0 J_c \to 0$, there is no technical noise from a current or a magnetic field source. However, the zero minimum of magnetic field magnitude in this trap has negative influence on the trapped atom cloud lifetime due to the Majorana effect (non-adiabatic spin-flips). Applying a bias field orthogonal to the film, it was possible (see [30]) to change the trap shape and also its position above the superconducting film (the zero minimum of the magnetic field magnitude remains).

For the magnetic trap generated under similar conditions above a thin superconducting disc, the problem is axisymmetric and the current density distribution is one-dimensional. Such a trap can be easily simulated numerically for the power law model (see [42]) or, for the Bean model, analytically using the known solution for this geometry [38]. Assuming the same two pulses of the external field, $0 \to 3\mu_0 J_c \to 0$ and $0 \to -0.8\mu_0 J_c \to 0$, were applied, we compare (figure 2), for different values of a normal to the film bias field, our simulation results for the 1 mm ×1 mm square film and for a disc of diameter 1 mm. For the square film we show (figure 2, left) two iso-surfaces of the magnetic field: $|\boldsymbol{B}| = 0.05\mu_0 J_c$ and $|\boldsymbol{B}| = 0.12\mu_0 J_c$. For the axisymmetric magnetic field above the disc, $|\boldsymbol{B}|$ is better represented by its cross-section contour plot (figure 2, right).

Without the bias field, the traps produced by these chips (figure 2, top) look similar: in both cases only one of these surfaces, $|\boldsymbol{B}| = 0.05\mu_0 J_c$ is closed and may represent a possible trap shape. The trap depth is 0.055 $\mu_0 J_c$ for the disc and about 0.07 $\mu_0 J_c$ for the square chip. A normal to the film magnetic field strongly influences the sheet current density in the film and also the trap depth, shape, and height above the superconductor. At a low bias field, $B_{bias} = 0.2\mu_0 J_c$, the possible trapping domain size increases, as well as the trap depth (up to 0.2 $\mu_0 J_c$ for the disc); the trap height decreases. The real shape of the atom cloud depends on the atom temperature: at a temperature corresponding to the low potential level $|\boldsymbol{B}| = 0.05\mu_0 J_c$ the shape is expected to be torus-like in both cases. At a higher temperature the atom cloud has, probably, no hole: its shape should be similar to that of the $|\boldsymbol{B}| = 0.12\mu_0 J_c$ iso-surface. The shape anisotropy of iso-surfaces becomes apparent in the square chip case.

Further increase of the bias field brings the trap closer to the film. For $B_{bias} = 0.4\mu_0 J_c$ the $|\boldsymbol{B}| = 0.12\mu_0 J_c$ iso-surfaces take torus-like shapes for both the square and disk chips. The $|\boldsymbol{B}| = 0.05\mu_0 J_c$ surface for the square chip splits into four separate closed surfaces, which suggests a possibility of splitting one atom cloud into four by increasing the bias field. The traps size and their distance from the chip surface decrease further for $B_{bias} = 0.6\mu_0 J_c$. According to our computation, in this case the $|\boldsymbol{B}| = 0.12\mu_0 J_c$ iso-surface is very close to the chip but, for the square chip, is still closed. For the circular chip such surface is not closed and touches the film. It should be noted though that the accuracy of our magnetic field calculation is lower very close to the conducting surface and, in addition, the finite film thickness should, probably, be taken into account in this case.

Trap splitting into four smaller separate traps, as well as the very close to the film surface position of these traps, have been reported in [30]. However, in our simulation the four traps are much closer to the square center. Under the specified conditions we found, near the square corners, no minima of the magnetic field magnitude which could correspond to the traps in [30]. We note that the shape and position of the closed $|\boldsymbol{B}|$ iso-surfaces strongly depend on the history of the magnetic field variations, including those during atom loading (not known to us exactly for this experiment).

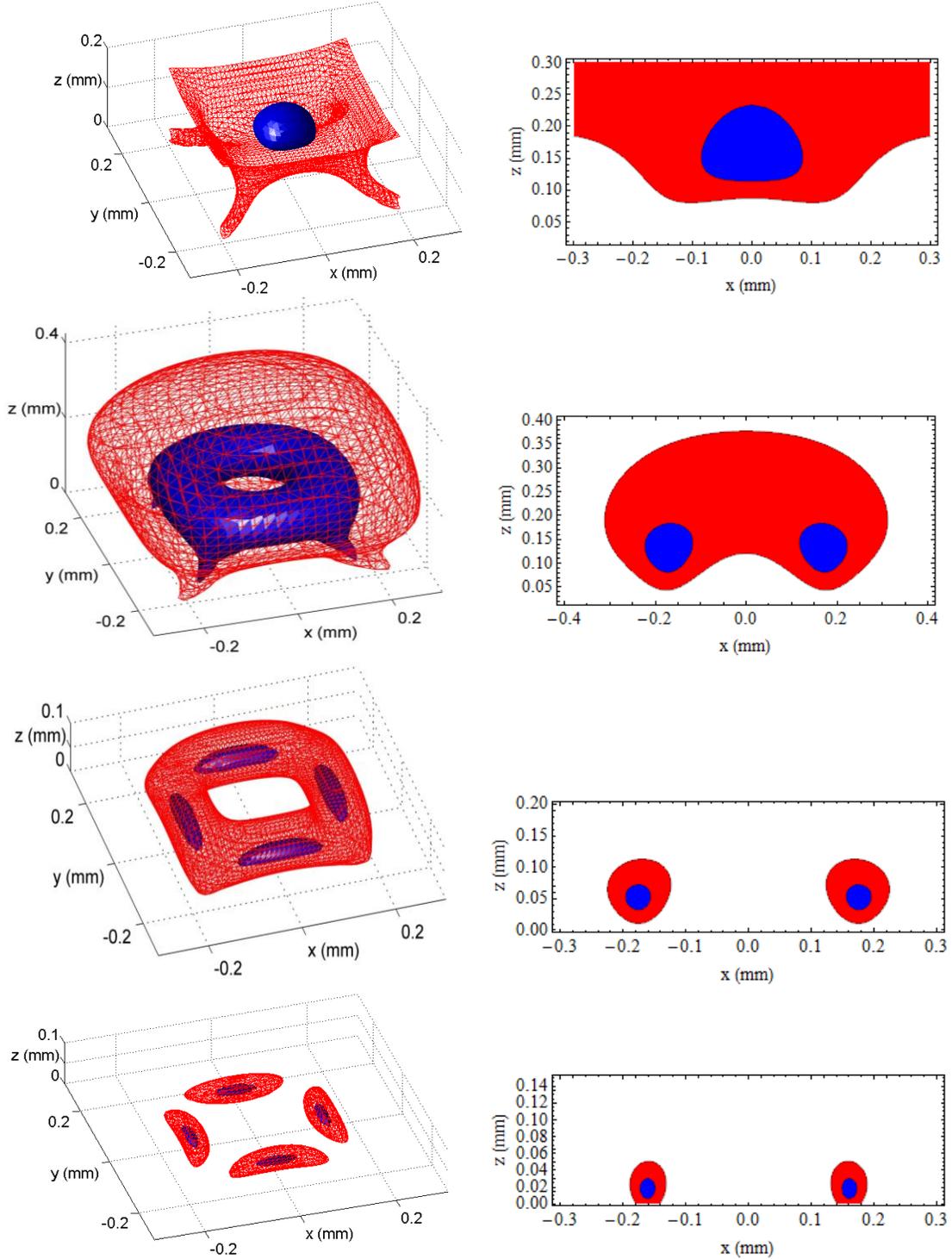

**Figure 2.** Simulations for magnetic traps for square (left) and circular (right) superconducting chips after two opposite pulses of external field, $0 \to 3\mu_0 J_c \to 0$ and $0 \to -0.8\mu_0 J_c \to 0$, supplemented by a normal to film bias field. The applied bias field $B_{bias}/\mu_0 J_c$ values are (from top to bottom): 0, 0.2, 0.4, 0.6. For the square chip two iso-surfaces are shown: $|\boldsymbol{B}|/\mu_0 J_c = 0.05$ (blue) and $|\boldsymbol{B}|/\mu_0 J_c = 0.12$ (red). For the circular chip we present a $|\boldsymbol{B}|/\mu_0 J_c$ contour plot (same levels) in the cross-section $y=0$.

## 3. A trap on a Z-shaped chip
### 3.1. Transport current

Magnetic traps on a Z-shaped superconducting film with a transport current have been created, e.g., in [14,19,24,27]. For this configuration the field induced by the film current should be supplemented by a bias field parallel to the film surface. In the infinitely thin film approximation, such a field does not change the sheet current density distribution. If no normal to the film external field was applied, the distribution is determined solely by the initial state and the history of applied transport current variations.

Let us consider a thin superconducting film consisting of two parallel to the *x*-axis long strips of width $w$ (the semi-infinite current leads) and a parallel to the *y*-axis central strip of length $l = 3w$ and width $w$ (figure 3). We assume the material properties of all film parts are similar. Initially, both the applied current and magnetic field are zero, and the film is cooled below the critical temperature. Then the transport current $I=0.7I_c$ is applied. The bias field $0.1\mu_0 J_c$ along the *x*-axis is also applied. Assuming the Bean critical-state current-voltage relation for the film we computed the resulting magnetic field as follows.

Sufficiently far from the central Z-shaped part of the film, the sheet current density distribution in the leads should be close to that in an infinite strip under the same conditions; the latter distribution is one-dimensional and known analytically for the Bean model [36,37] (for the power law model it can be easily calculated numerically). It was sufficient to cut off the leads at the distance $3w$ from the central strip and to use the distribution of current in the infinite strip as a boundary condition for the sheet current density on the cuts (see Appendix). To fully account for the cut-off semi-infinite lead parts, we also added the "external" magnetic field induced by their current, assumed equal to that in the infinite strip.

The computational domain (the remaining part of the film) becomes bounded and the sheet current density in this domain was computed for the Bean model (approximated by the power law model with $p = 10^6$) on a finite element mesh of about thirty thousand triangles. For such a mesh the computations are time and, especially, memory consuming (see [49]); they were performed on a 64 GB RAM, 2.0 HGz Intel(R) Xeon E5-2620 2 computer with 6 CPUs. Computing the current density took several hours; then the magnetic field was calculated.

The closed iso-surfaces $|\mathbf{B}| = 0.065\mu_0 J_c$ and $|\mathbf{B}| = 0.04\mu_0 J_c$ in figure 3 are presented as possible shapes of a 3D magnetic trap on the Z-shaped superconducting chip. The minimum of magnetic field magnitude inside this trap is found at the height $0.97w$ and is nonzero (about $0.021\mu_0 J_c$) due to the field induced by the lead currents. This field, significant because the chip central strip is not long, decreases the Majorana instability and makes atom trapping more stable without any additional field.

We note that to model such, and similar, trap configurations, Z- and U-shaped chips are sometimes replaced by a straight infinite strip [9,41] for which calculating the induced magnetic field is easier; this field is used to approximate the cross section of a magnetic trap in its central part. Such an approach provides no information about the trap ends. Moreover, it is applicable only if the length $l$ of the central strip is much greater than the strip width $w$ and the magnetic field of the lead currents is negligible near the central part of this strip. For a straight infinite strip the minimum of the magnetic field magnitude in the trap is zero, in contrast to the Z-chip case considered above. Our simulation showed also that, for the same transport current and bias field, the $|\boldsymbol{B}|$ iso-surfaces differ significantly from those for the straight strip (figure 3, bottom); the simplified model can, therefore, be inaccurate.

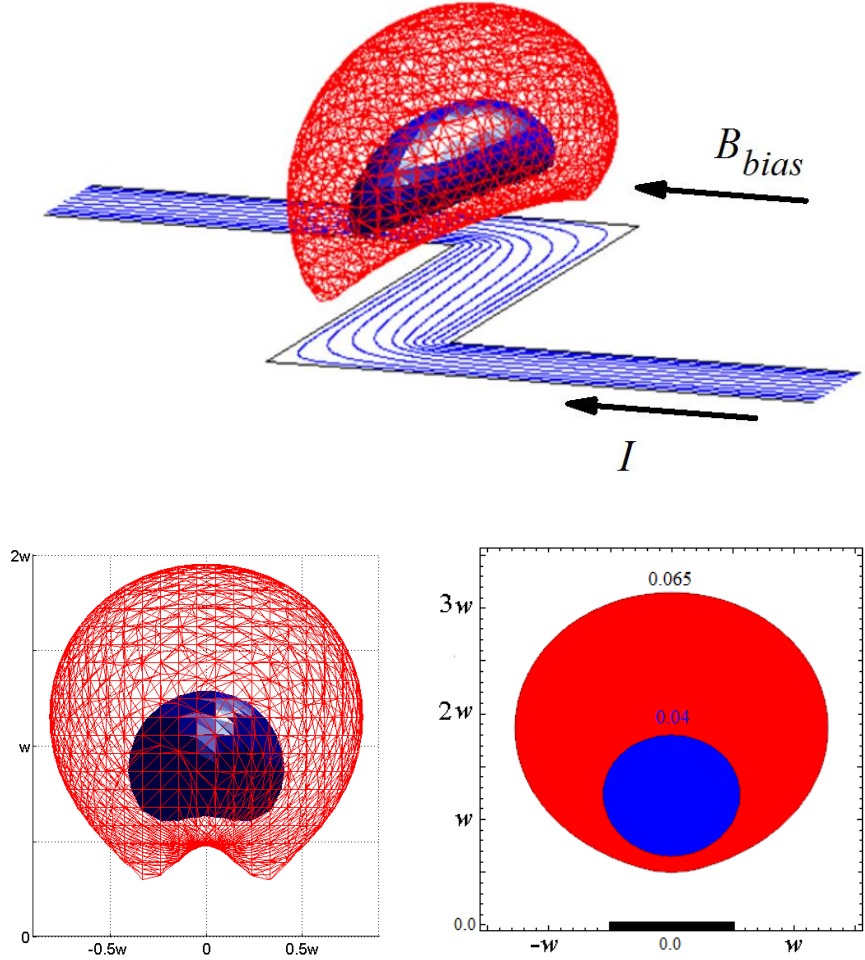

**Figure 3**. A trap on a Z-shaped superconducting chip. The transport current $I=0.7I_c$ and the bias field $B_{bias}=0.1\mu_0 J_c$ are applied. Shown: the level surfaces $|\boldsymbol{B}|=0.065\mu_0 J_c$ (red) and $|\boldsymbol{B}|=0.04\mu_0 J_c$ (blue). Top: an isometric view; the current streamlines are also shown (blue lines). Bottom left: the end view of the trap. Bottom right: cross-section $|\boldsymbol{B}|$ contour plot for the infinitely long strip.

### 3.2. Transport current pulse

In our last example, the transport current pulse $0\rightarrow 0.7I_c \rightarrow 0$ is applied. Even though the transport current in the Z-shaped superconducting film returns to zero, a sheet current density with a non-trivial 2D distribution remains (figure 4, blue lines). In this case some of the current stream lines are closed in the vicinity of the film corners; the rest of the lines are closed far away where the leads are disconnected. The magnetic field, induced by this current, is again supplemented by a bias field parallel to the $x$-axis. The induced field is now much weaker than in the previous example, and the applied bias field should be weaker as well. The same mesh as above was used to compute the sheet current density and then the resulting magnetic field in a vicinity of the film.

We present (figure 4) the $|\mathbf{B}| = 0.004\mu_0 J_c$ iso-surfaces for different bias fields to show how this field controls the possible trap shape. A single closed trap at $B_{bias} = 0.006\mu_0 J_c$ splits for $B_{bias} = 0.008\mu_0 J_c$ into three traps separated by a potential barrier. Further increase of the bias field ($B_{bias} = 0.012\mu_0 J_c$) causes disappearance of the small trap in the middle.

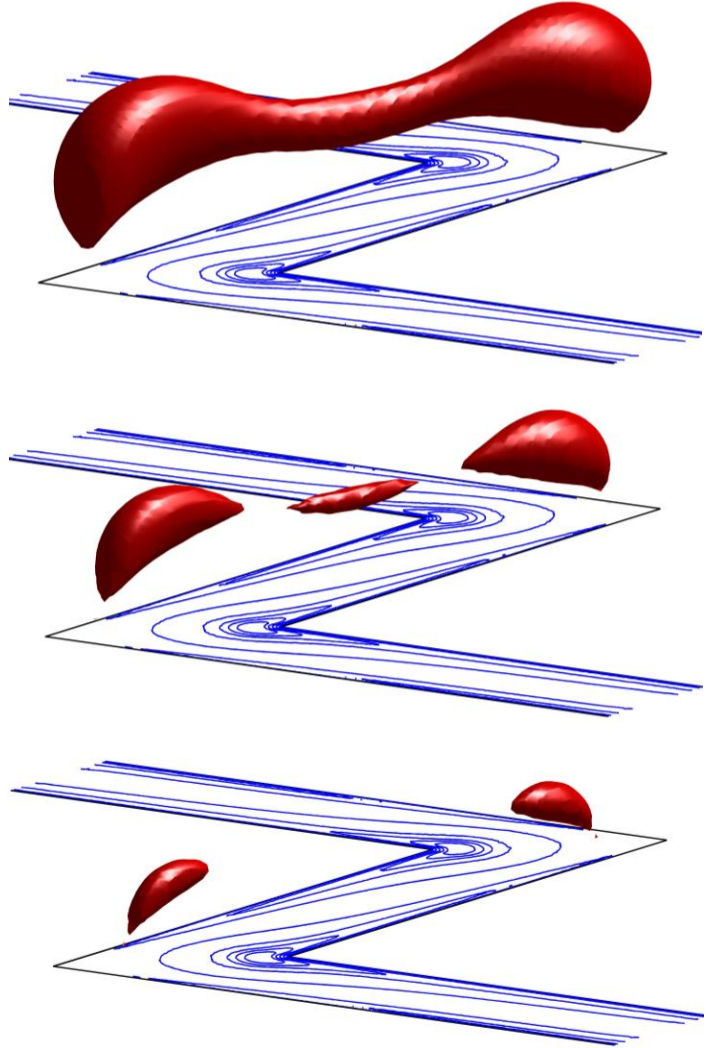

**Figure 4**. A trap on a Z-shaped superconducting chip after the transport current pulse $0 \to 0.7 I_c \to 0$. Influence of the bias field on the magnetic trap above a Z-shaped film. The magnetic field iso-surface $|\mathbf{B}| = 0.004\mu_0 J_c$ is shown (from top to bottom) for $B_{bias} / \mu_0 J_c = 0.006, 0.008, 0.012$. The current stream lines are also shown (the blue lines).

## 4. Discussion

In the Bean critical-state model, the evolution of film sheet current density, driven by temporal variations of the external magnetic field and transport current, is rate-independent. The normalized magnetic field, induced by the film current, can be presented in the following form,

$$\frac{\boldsymbol{B}_{ind}}{\mu_0 J_c} = \boldsymbol{\Phi}\left[\frac{\boldsymbol{r}}{w}, \frac{B_{ext}}{\mu_0 J_c}, \frac{I}{J_c w}\right],$$

where $\boldsymbol{r}=(x,y,z)$ is the radius vector and the operator $\boldsymbol{\Phi}$ takes into account the zero initial condition and variation history of both the transport current $I$ and the normal to film component of the external magnetic field, $B_{ext}$.

For a uniform external field the normalized magnetic field $\boldsymbol{B}_{ind}/\mu_0 J_c$ is independent of the film size $w$. The parallel to film bias external magnetic field, applied to change and control the magnetic trap, is chosen in accordance with $\boldsymbol{B}_{ind}$. Therefore, in dimensional variables, the trap depth is proportional to $\mu_0 J_c$ and, for similar film shapes and the same $I/J_c w$ ratio, does not depend on the film size $w$. The size of a trap is, on the contrary, proportional to $w$. The magnetic field gradient, determining the steepness of the potential well in a trap, is, therefore, proportional to $\mu_0 J_c/w$. These similarity arguments are important for planning the atom trapping experiments and designing a superconductor chip.

The main characteristic of thin superconducting films is their critical sheet current density $J_c$. This density depends on the superconductor material, substrate, temperature, fabrication technology, etc.; see e.g. [51,52,53]. In atom trap experiments, the niobium (Nb) [14,15,19,27,54,55], magnesium diboride ($MgB_2$) [25,26], and high-temperature YBCO [28,29,56] films have been employed.

The critical temperature of Nb films is about 9.5 K; usually, their thickness is within the range 400÷900 nm and the chip operation temperature is 4-6 K [14,15,19,27,54,55]. Under such conditions the critical sheet current density $J_c$ of Nb films is in the range $(1.6 \div 3.6) \times 10^4$ A/m. The characteristic magnetic field $B_{ch} = \mu_0 J_c$ is, therefore, between 200 G and 450 G.

Although the critical temperature of $MgB_2$ superconductors is much higher, up to 40 K, the $MgB_2$ chips are also cooled to 4 K [25,26] in order to increase the critical density $J_c$ to $1.6 \times 10^5$ A/m at the film thickness $d_f = 1.6$ μm; the corresponding characteristic field $B_{ch}$ is estimated as 2000 G.

High-temperature YBCO superconducting films with the critical temperature of ~90 K allow one to use liquid nitrogen at 77 K for cooling. It is known that, usually, the sheet critical current density of YBCO films is not proportional to their thickness. At 77 K for $d_f$ ~300 nm the typical sheet critical value is $1.1 \times 10^4$ A/m; for $d_f = 600 \div 800$ nm, $J_c = (1.2 \div 2.1) \times 10^4$ A/m [29,56]. Respectively, $B_{ch}$ equals to 140 G or is in the range 150÷260 G. In some experiments the superconductor temperature was higher, 83 K [28,29], and the sheet critical current density decreased to $0.4 \times 10^4$ A/m for $d_f = 300$ nm. The Ag doping of a multilayered film structure [52] resulted in the sheet critical current density $J_c = 3 \times 10^4$ A/m at 77 K and $J_c = 30 \times 10^4$ A/m at 10 K for $d_f = 1$ μm. These critical values correspond to the characteristic magnetic fields $B_{ch}$ equal to 380 G and 3800 G, respectively.

We conclude that, at low temperatures, 4-10 K, the MgB$_2$ and YBCO films enable one to create a much deeper trap than the Nb chips. At higher temperatures (77 K and above), the trap depth of YBCO chips is comparable to that of Nb chips at low temperature.

Using the stability criterion $\mu B_{dep} \geq 10 k_B T$ and estimating the force, acting in an inhomogeneous magnetic field on the most often employed in experiments $^{87}$Rb atoms in the $|F=2, m_F=2\rangle$ state, it was found [9] that at the atom gas temperature 1 µK the trap depth should be not less than 0.07 G and, to protect the atoms from gravity's pull, the field gradient should be at least 15 G/cm (here $F$ is the total atom spin and $m_f$ its projection on the local field).

Numerical simulations enable us to calculate the depth of magnetic traps on superconducting chips and to estimate the trapping field gradient. The Z-shaped MgB$_2$ film in [26,27] had the width $w$ =100 µm and thickness 1.6 µm. Assuming $\mu_0 J_c$ is about 2000 G, we use as an example the traps modeled in dimensionless form for a Z-shaped chip in Section 3.

In the case of the transport current $0 = 0.7 I_c$ and the bias field $0.1 \mu_0 J_c$ (figure 3), the depth of a trap represented by the $|\mathbf{B}|= 0.065 \mu_0 J_c$ iso-surface is $0.044 \mu_0 J_c$ =88 G and, correspondingly, the magnetic field gradient is of the order of $0.044 \mu_0 J_c / w$ = 8800 G/cm.

Simulations for the transport current pulse $0 \to 0.7 I_c \to 0$ yield that, e.g., if the bias field is $0.012 \mu_0 J_c$, the two traps represented by the $|\mathbf{B}|=0.004 \mu_0 J_c$ iso-surface (figure 4, bottom) have the depth, approximately, $0.0035 \mu_0 J_c = 7$ G, and the field gradient is of the order of 700 G/cm. Hence, even though the magnetic field is much weaker in the latter case, it should be sufficient to trap $^{87}$Rb atoms at 1 µK.

Varying the bias field, one can split a trap into several traps (figure 2 and 4) and merge them into one trap again. For the Z-shaped chip trap (figure 4) having a nonzero minimum of the magnetic field magnitude such a behavior can probably be used for experimental investigation of coherence and decoherence of atom clouds, tunneling of cold atoms, including atoms in the Bose-Einstein condensate state, and to build an atom interferometer in the temporal domain.

We believe that 3D magneto-dynamic modeling of atom traps on superconducting chips, as in our work, helps to design and analyze magnetic traps for cold atoms. Our model takes into account general dependence of the trap properties on the atom cloud temperature, the film shape, and the history of applied currents and magnetic fields. However, while we assume the trap shape can be approximated by a closed $|\mathbf{B}|$ iso-surface, the density of atom distribution in a trap, observed in experiments, is not simulated in our work. Among other factors, not included into our model but able to affect the shape of the atom cloud, are: gravitational and other possible forces, inhomogeneity of superconductors, the finite thickness of superconducting films, etc. In future simulations these factors should possibly be taken into account.

5. **Conclusion**

This work presents an approach to 3D modeling of magnetic atom traps based on superconducting chips. The main chip element, a flat thin superconducting film in the mixed state, can be of an arbitrary shape.

Using the finite element method [49], based on an evolutionary variational formulation of thin film magnetization and transport current problems in type-II superconductivity, we first compute the 2D sheet current density distribution in the film. The method is applicable for both the power and the critical-state current voltage relations characterizing the superconducting material. Then the 3D magnetic field, induced by the film current, is found numerically by an accurate integration [21] of the Biot-Savart law. Finally, the trap shape is represented by a closed iso-surface of the total magnetic field magnitude; the level is chosen in accordance with the atom cloud temperature.

Our simulations have been performed for the chip configurations employed in recent cold atom experiments. The developed approach takes into account the superconductor properties and the variation of the external magnetic field and transport current and enables one to analyze such important characteristics of the magnetic traps as their depth, size, shape, and distance from the chip surface. Knowledge of these characteristics is important for designing a cold atom physics experiment.

**Acknowledgment**
The authors thank R. Folman for helpful discussions and comments.

**Appendix**. **Variational formulation and numerical solution**
Let, in the infinitely thin film approximation, the flat film lie in the plane $z = 0$ and the transport current $I(t)$ be supplied to the film by means of two semi-infinite superconducting strip leads of the width $w$ (figure A1) lying in the same plane.

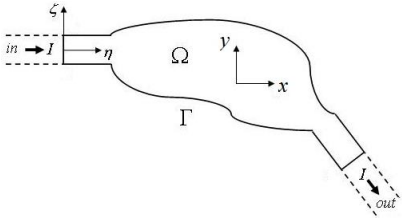

**Figure A1**. Thin film with a transport current; dashed lines show the cut-off lead ends.

We assume the film and leads are made of the same superconducting material, characterized by the power current voltage relation

$$\boldsymbol{E} = E_0 \left(\frac{|\boldsymbol{J}|}{J_c}\right)^{p-1} \frac{\boldsymbol{J}}{J_c} \quad \text{or} \quad \boldsymbol{J} = J_c \left(\frac{|\boldsymbol{E}|}{E_0}\right)^{q-1} \frac{\boldsymbol{E}}{E_0}, \tag{1}$$

where $\boldsymbol{E}$ is the parallel-to-film component of the electric field, $\boldsymbol{J}$ is the sheet current density, $E_0$, $p$, and $J_c$ are constants, and $q = 1/p$. By $H_z^e(t)$ we denote the normal-to-film component of a given uniform external magnetic field. Sufficiently far from the film, the sheet current density in the leads should be close to that in an infinite strip carrying the transport current $I(t)$ and exposed to the field $H_z^e(t)$. This 1D current density distribution, $\tilde{\boldsymbol{J}} = \left(\tilde{J}_\eta(\zeta,t),0\right)$ in the strip-related coordinates $\{\eta,\zeta\}$ (see figure A1), can be found numerically for the power law model and analytically for the Bean model [35-37,45]. Let us cut off the semi-infinite leads at a sufficient distance from the film and set the normal to the boundary component of the sheet current density $J_n = \pm\tilde{J}_\eta$ on the out- and in- lead cuts, respectively, and $J_n = 0$ on the other part of the boundary $\Gamma$ of the remaining film part $\Omega$. We also add to $H_z^e(t)$ the normal to the film component of the magnetic field induced by the current in the cut-off lead ends,

$$\tilde{H}_z^e = H_z^e(t) + H^{in} + H^{out},$$

where (see [21,49])

$$H^{in}(\eta,\zeta,t) = \frac{1}{4\pi} \int_{-w/2}^{w/2} \frac{\tilde{J}_\eta(\zeta',t)}{\zeta-\zeta'} \left[1 - \frac{\eta}{\sqrt{(\zeta-\zeta')^2 + \eta^2}}\right] d\zeta' \quad (2)$$

for $\eta > 0$; similarly for $H^{out}$. Following [49], we now substitute the Biot-Savart law

$$H_z = \tilde{H}_z^e + \frac{1}{4\pi} \int_\Omega \mathbf{Curl'}\left(\frac{1}{|\boldsymbol{r}-\boldsymbol{r'}|}\right) \cdot \boldsymbol{J}(\boldsymbol{r'},t) d\boldsymbol{r'},$$

where $\boldsymbol{r} = (x,y) \in \Omega$, into the Faraday law $\mu_0 \partial_t H_z + \text{Curl}\, \boldsymbol{E} = 0$ and obtain

$$\partial_t \tilde{H}_z^e + \frac{1}{4\pi} \int_\Omega \mathbf{Curl'}\left(\frac{1}{|\boldsymbol{r}-\boldsymbol{r'}|}\right) \cdot \partial_t \boldsymbol{J}(\boldsymbol{r'},t) d\boldsymbol{r'} + \frac{1}{\mu_0} \text{Curl}\, \boldsymbol{E} = 0.$$

Here $\text{Curl}\,\boldsymbol{f} := \partial_x f_y - \partial_y f_x$ and $\mathbf{Curl}\, u(x) := (\partial_y u, -\partial_x u)$ are 2D operators.

Since $\text{Div}\,\boldsymbol{J} = 0$, we can introduce the stream function $g$ such that $\boldsymbol{J} = \mathbf{Curl}\, g$ in the domain $\Omega$ and, on its boundary, $g(\boldsymbol{r},t) = \int_{\Gamma(\boldsymbol{r}_0,\boldsymbol{r})} J_n ds$, where the integration from a fixed point $\boldsymbol{r}_0 \in \Gamma$ to the point $\boldsymbol{r} \in \Gamma$ is in the counter-clockwise direction along the boundary. It is convenient to use the transformation $\boldsymbol{V} = (-E_y, E_x)$ and rewrite the equations (1)-(2) in terms of the new variables, $g$ and $\boldsymbol{V}$, as

$$\mathbf{Grad}\, g = -J_c \left(\frac{|\boldsymbol{V}|}{E_0}\right)^{q-1} \frac{\boldsymbol{V}}{E_0} \quad (3)$$

and

$$a(\partial_t g, \varphi) - \frac{1}{\mu_0}(\boldsymbol{V}, \mathbf{Grad}\, \phi) = -\left(\partial_t \tilde{H}_z^e, \phi\right) \quad (4)$$

for any smooth enough test function $\phi$ which is zero on $\Gamma$. Here

$$(\boldsymbol{U},\boldsymbol{V}) = \int_\Omega \boldsymbol{U}(\boldsymbol{r}) \cdot \boldsymbol{V}(\boldsymbol{r}) d\boldsymbol{r}, \quad a(\psi,\varphi) = \frac{1}{4\pi} \iint_{\Omega\Omega} \frac{\mathbf{Grad}\,\psi(\boldsymbol{r}) \cdot \mathbf{Grad'}\,\phi(\boldsymbol{r'})}{|\boldsymbol{r}-\boldsymbol{r'}|} d\boldsymbol{r}\, d\boldsymbol{r'}$$

and, to complete the model, one should add the boundary and an initial condition for the stream function $g$. We note that for the magnetization problems (with zero transport current and no leads) the boundary condition is simply $g|_\Gamma = 0$ and also $\tilde{H}_z^e = H_z^e$.

To solve the problem (3)-(4) numerically, we employed (see [49]) an implicit discretization of the variational equation (4) in time and, for the approximation in space, triangulated $\Omega$ and used the non-conforming linear and the piecewise constant finite elements for $g$ and $V$, respectively. The iterations, needed on each time level to deal with the nonlinearity in (3), were based on approximating the term $|V^k|^{q-1} V^k$ by

$$|V^{k,j-1}|^{q-1} V^{k,j-1} + \left(|V^{k,j-1}|_\varepsilon\right)^{q-1}\left(V^{k,j} - V^{k,j-1}\right),$$

where $k$ and $j$ are the time level and iteration numbers, respectively, and $|V|_\varepsilon = \sqrt{|V|^2 + \varepsilon^2}$ with a small $\varepsilon$ (in our numerical examples $\varepsilon = 10^{-10}$). It was possible to accelerate these iterations by using an over-relaxation algorithm. Convergence of this numerical method to a solution of (3)-(4) was proved for the magnetization problems [57].

Nonconforming linear approximation of the stream function $g$ is a function, linear on each mesh triangle and continuous at the midpoints of triangle edges. We found that the piecewise constant approximation of the sheet current density $J^k$, computed in each triangular element $\sigma$ directly as $J^k|_\sigma = \mathbf{Curl}\left(g^k|_\sigma\right)$, can be inaccurate in problems with a transport current. In these problems it was desirable to approximate first the obtained piecewise linear but discontinuous function $g^n$ by a continuous piecewise linear function $\hat{g}^n$, then to calculate $J^k|_\sigma = \mathbf{Curl}\left(\hat{g}^k|_\sigma\right)$ (see [49]). Finally, we computed the magnetic field in the vicinity of the film by integrating numerically (see [21]) the Biot-Savart law for this approximation of the sheet current density and adding the applied external field and the field of the current in the cut-off lead ends (also calculated numerically).